\documentclass[%
amsmath,amssymb,prl,superscriptaddress,twocolumn
]{revtex4-2}

\usepackage{graphicx}
\usepackage{dcolumn}
\usepackage{bm}
\usepackage{hyperref}
\usepackage{natbib}
\usepackage{float}
\usepackage{amsmath}
\usepackage{xcolor}
\usepackage{braket}

\newcommand{\beginsupplement}{
    \onecolumngrid
    \setcounter{page}{1}
    \setcounter{table}{0}
    \renewcommand{\thetable}{S\arabic{table}}
    \setcounter{figure}{0}
    \renewcommand{\thefigure}{S\arabic{figure}}
    \setcounter{equation}{0}
    \renewcommand{\theequation}{S\arabic{equation}}
    \setcounter{secnumdepth}{2}
}

\begin{document}

\title{Spin splitting, Kondo correlation and singlet-doublet quantum phase transition in a superconductor-coupled InSb nanosheet quantum dot}

\author{Xingjun Wu}
\email{wuxj@baqis.ac.cn}
\affiliation{Beijing Academy of Quantum Information Sciences, Beijing 100193, China}

\author{Ji-Yin Wang}
\affiliation{Beijing Academy of Quantum Information Sciences, Beijing 100193, China}
\author{Haitian Su}
\affiliation{Beijing Academy of Quantum Information Sciences, Beijing 100193, China}
\author{Han Gao}
\affiliation{Beijing Academy of Quantum Information Sciences, Beijing 100193, China}
\author{Shili Yan}
\affiliation{Beijing Academy of Quantum Information Sciences, Beijing 100193, China}
\author{Dong Pan}
\email{pandong@semi.ac.cn}
\affiliation{State Key Laboratory of Semiconductor Physics and Chip Technologies, Institute of Semiconductors,Chinese Academy of Sciences, P.O. Box 912, Beijing 100083, China}

\author{Jianhua Zhao}
\affiliation{State Key Laboratory of Semiconductor Physics and Chip Technologies, Institute of Semiconductors,Chinese Academy of Sciences, P.O. Box 912, Beijing 100083, China}
\affiliation{National Key Laboratory of Spintronics, Hangzhou International Innovation Institute, Beihang University, Hangzhou 311115, China}
\author{Po Zhang}
\affiliation{Beijing Academy of Quantum Information Sciences, Beijing 100193, China}

\author{H. Q. Xu}
\email{hqxu@pku.edu.cn}
\affiliation{Beijing Academy of Quantum Information Sciences, Beijing 100193, China}
\affiliation{Beijing Key Laboratory of Quantum Devices, Peking University, Beijing 100871, China}

\begin{abstract}
We realize a superconductor-coupled quantum dot (QD) in an InSb nanosheet, a 2D platform promising for studies of topological superconductivity. The device consists of a superconductor-QD-superconductor junction, where a bottom bilayer gate defines the QD and allows tuning of its coupling to the superconducting leads. The QD exhibits large $g$-factors and strong spin-orbit coupling. Transport measurements reveal Coulomb diamond-shaped differential conductance features with even-odd alternating sizes and pronounced conductance lines associated with the superconducting gap, confirming a few-electron, superconductor-coupled regime. At an odd electron occupation, Kondo signatures emerge, including a zero-bias peak that splits with magnetic field and is logarithmically suppressed at elevated temperatures. We further observe a doublet-singlet quantum phase transition, manifested by a clear change of Andreev bound states from crossing to anticrossing as the coupling strength increases. These results underscore the rich physics of InSb nanosheet QDs and their promise for topological quantum devices.
\end{abstract}

\maketitle

Quantum dots coupled to superconductors in materials with strong spin-orbit coupling are pivotal for the development of topological superconducting quantum computing chips. They enable the realization of Majorana zero modes in artificial Kitaev chains~\cite{leijnse2012parity,sau2012realizing,fulga2013adaptive,dvir2023realization,ten2024two,bordin2025enhanced} and serve as critical components for the readout and manipulation of topological qubits~\cite{karzig2017scalable,plugge2017majorana,microsoft2025interferometric}. Among these, InSb QDs have emerged as a prime candidate for engineering topological superconductivity and quantum devices, owing to the material's unique properties, including a small electron effective mass, high mobility, a large $g$-factor, and strong spin-orbit coupling. Extensive studies have focused on InSb nanowire QDs~\cite{nilsson2009giant,nilsson2010correlation,nadj2012spectroscopy,deng2012anomalous,pribiag2013electrical,deng2014parity,fan2015formation,li20170,pendharkar2021parity,dvir2023realization,deng2024anomalous,bordin2025enhanced,deng2025quantum}, as one-dimensional (1D) structures facilitate QD confinement via electrical gating. In contrast, 2D architectures offer additional dimensional degrees of freedom, enabling the complex geometries required for topological quantum computing~\cite{microsoft2025interferometric,ten2024two,stern2019fractional,hell2017coupling,hell2017two}. They further provide enhanced versatility for topological applications---specifically, via phase control of superconductors in planar architectures---and thus hold significance for studies of phase-tuned topological superconductivity~\cite{hell2017two,pientka2017topological,fornieri2019evidence} and multiterminal Josephson junctions~\cite{riwar2016multi,coraiola2024spin,gupta2023gate,matsuo2023josephson}.

\par Recent progress in 2D InSb systems has been remarkable. The techniques for growth of high-quality InSb quantum wells and nanosheets have advanced substantially~\cite{yi2015gate,de2016twin,pan2016free,lehner2018limiting,lei2022high}, and strong, gate-tunable spin–orbit coupling in these 2D materials has been experimentally demonstrated~\cite{chen2021strong,lei2023gate}. In 2D InSb-based Josephson junctions, ballistic transport has been observed~\cite{ke2019ballistic,salimian2021gate}, and proximity-induced superconductivity has been extensively explored~\cite{kang2018two,zhi2019coexistence,turini2022josephson,iorio2023half,chieppa2025unveiling,wu2025tunable,3tvf-5v29}. Furthermore, recent demonstrations of gate-defined nanoconstriction structures, including quantum point contacts and quantum dots~\cite{xue2019gate,kulesh2020quantum,lei2021gate,chen2021double}, have provided a strong impetus for investigating superconductor-coupled QDs defined in planar InSb structures. However, the experimental realization of a planar InSb QD coupled to superconductors has yet to be reported, hindered by material- and fabrication-related challenges~\cite{lei2024quantum}. 

\par In this work, we report the realization of a QD in an InSb nanosheet using a predefined bilayer fine-gate structure on the substrate. After the InSb nanosheet was transferred onto the predefined bilayer gate structure, we fabricated superconducting electrodes atop the nanosheet via a simple process, enabling the realization of a superconductor-coupled QD on a 2D InSb nanosheet in a single step. Since no multi-step processing was required on the InSb nanosheet, potential material degradation and deterioration of the interface between InSb and the superconductor, which can be induced during additional fabrication processes, were avoided. The fabricated device was studied by transport spectroscopy measurements. The results reveal a pronounced Coulomb blockade effect, with an even-odd alternation in Coulomb diamond size, along with a large $g$-factor and strong spin-orbit coupling, similar to the behavior observed in their 1D nanowire counterparts~\cite{nilsson2009giant}. In a Coulomb diamond with an odd-number electron occupation, we also observed clear Kondo signatures, including a zero-field Kondo ridge that splits as the magnetic field increases and a logarithmic suppression of the conductance peak value at elevated temperatures. Temperature-dependent measurements reveal a higher Kondo temperature $T_{K}$ at energy levels closer to a diamond’s charge degeneracy point. Furthermore, modulating the superconductor-QD coupling strength reveals a transition of Andreev bound states (ABSs) from crossing to anticrossing characteristics, directly associated with the doublet-singlet quantum phase transition in the QD Josephson junction.

\begin{figure*}
\includegraphics[width=0.5\linewidth]{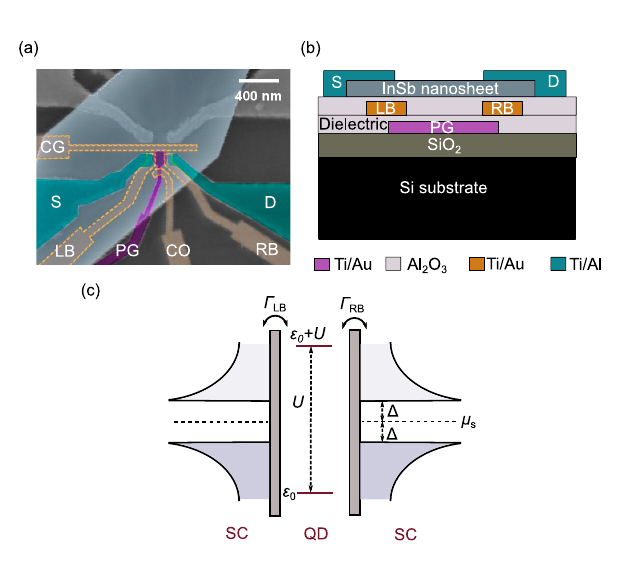}
\caption{ 
(a) False-colored scanning electron microscopy image of a QD in an InSb nanosheet. Two lithographically defined metallic gate layers, indicated by pink and orange dashed lines, lie beneath the nanosheet and are separated by $\text{Al}_{2}\text{O}_{3}$ dielectric. These gates are used to define a QD in the InSb nanosheet. The green superconducting contacts (S, D), positioned atop the InSb nanosheet, act as the source and drain electrodes, respectively. (b) Cross-sectional schematic of the layer structure of the nanosheet QD device. (c) Schematic energy diagram of a QD Josephson junction device as studied in (a). $\Delta$ is the superconducting gap, $\varepsilon_{0}$ represents the QD energy level relative to the chemical potential of the right SC $\mu_{s}$, $U$ is the QD charging energy, and $\Gamma_{LB}$ ($\Gamma_{RB}$) is the coupling strength of the QD with left (right) superconducting contact.}
\label{fig1}
\end{figure*}

High-quality InSb nanosheets are grown by molecular beam epitaxy (MBE)~\cite{pan2016free}. Details on material and transport properties are provided in our previous reports~\cite{kang2018two,wu2025tunable,3tvf-5v29}. A false-colored scanning electron micrograph of the studied device is presented in Fig.~\ref{fig1}a and its corresponding cross-sectional view of the gate stack is illustrated in Fig.~\ref{fig1}b. The InSb nanosheet was deposited on a Si/SiO$_{2}$ substrate patterned with prefabricated bilayer metallic electrostatic gates (lower-layer gates marked in pink and upper-layer ones in orange). Each gate consists of 2.5-nm-thick Ti and 6.5-nm-thick Au, and was capped with a 10-nm-thick Al$_2$O$_3$ film grown via atomic layer deposition. Two Ti/Al superconducting contacts (SCs), 125~nm wide each and separated by 150 nm, were fabricated on the InSb nanosheet using standard electron-beam lithography after removing the native surface oxide. In this work, the QD is defined using a bilayer electrostatic gate architecture. While a single-layer gate design may be sufficient for defining an InSb QD~\cite{kulesh2020quantum}, a bilayer gate geometry provides enhanced tunability of the dot size and electron occupation, facilitating access to the few-electron regime. Specifically, the bilayer electrostatic gates serve the following functions: the plunger gate (PG) tunes the electrochemical potential ($\varepsilon_{0}$) of the QD; the left and right barrier gates (LB, RB) modulate tunnel couplings between the QD and SCs ($\Gamma_{\text{LB}}$ and $\Gamma_{\text{RB}}$, respectively); and the control gate (CG) and cutoff gate (CO) supply tuning the QD confinement. Gates and superconductor electrodes in an identical configuration, obscured by the nanosheet in the image, are symmetrically positioned about the control gate, but not employed in this work. To minimize their potential influences on the investigated superconductor-coupled QD, these gates are biased at -0.5 V, while the superconductor contacts remain floating. All measurements were conducted in a dilution refrigerator with a base temperature of 25 mK, where devices were vertically mounted in a plane parallel to the applied magnetic field.

\par
\begin{figure*}
\includegraphics[width=0.5\linewidth]{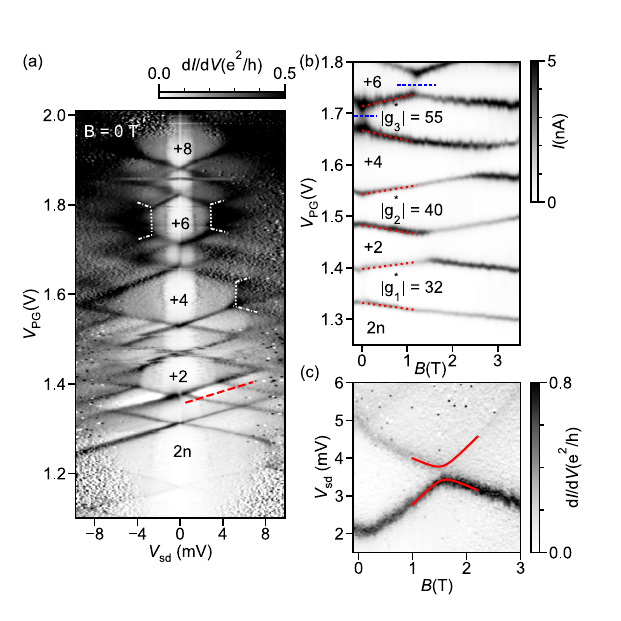}
\caption{
(a) Differential conductance d$I$/d$V$ as a function of source-drain bias $V_{\text{sd}}$ and plunger gate voltage $V_{\text{PG}}$ at $T$ = 25~mK and $B$ = 0~T, showing Coulomb diamonds typical for the electron transport in a few-electron QD. Several white dotted lines have been superimposed to illustrate the onset of inelastic cotunneling. The dash-dotted lines denote the onset of tunneling through an excited state. (b) Source-drain current as a function of the plunger gate voltage $V_{\text{PG}}$ and magnetic field $B$ at $V_{\text{sd}}$ = 1.2~mV. Conductance enhancements are observed in the Coulomb blockade regime along the blue dashed lines. (c) Differential conductance d$I$/d$V$ along the red cut in (a) as a function of magnetic field $B$. The red lines denote fits from a two-level model incorporating the spin-orbit energy $\Delta_{\text{SO}}$.
}
\label{fig2}
\end{figure*}

 To define a QD in the InSb nanosheet, a positive voltage $V_{\text{PG}}$ is applied to the plunger gate to accumulate electrons and form the dot, while small negative voltages, applied to the left barrier, right barrier, cutoff and control gates ($V_{\text{LB}}$, $V_{\text{RB}}$, $V_{\text{CO}}$, $V_{\text{CG}}$), create confining potential barriers. Figure~\ref{fig2}a shows the differential conductance, d$I_{\text{sd}}$/d$V_{\text{sd}}$, as a function of plunger gate voltage $V_{\text{PG}}$ and source-drain voltage $V_{\text{sd}}$ at gate voltages ($V_{\text{LB}}$, $V_{\text{RB}}$, $V_{\text{CO}}$, $V_{\text{CG}}$) = (-0.43, -0.6, -0.18, 0) V.  The observed diamond-shaped features exhibit a clear even-odd alternation in size---a hallmark of electron transport through a few-electron QD in the Coulomb blockade regime. The average voltage-to-energy conversion factor (lever arm) $\alpha$ = 0.08 for $V_{\text{PG}}$ is extracted from the slopes of the diamond edges. From the lateral dimensions of smaller Coulomb diamonds, we further deduce an average charging energy $U$ = 4.4~meV, corresponding to an averaged total QD capacitance $C_{\Sigma}$ = 36 aF. In addition, we observe multiple conductance enhancements inside the Coulomb diamonds at finite source–drain biases. Two high-conductance lines or edges, located symmetrically around zero bias, exhibit monotonic shifts in their $V_{\text{sd}}$ positions as $V_{\text{PG}}$ decreases, while other lines appear at higher biases and connect to the QD’s excited-state conductance lines [see the white lines in Fig. 2(a)]. We attribute the former probably to the activation of additional transport channels in the lead regions of the nanosheet, and the latter to the onset of inelastic cotunneling~\cite{de2001electron}.

\par InSb nanowire QDs exhibit strong spin-orbit coupling and large $g$-factors~\cite{nilsson2009giant}, both of which are critical for realizing topological superconductivity. To determine the $g$-factors of the nanosheet QD, we measure the magnetic-field dependence of Coulomb blockade peaks in the linear response regime. Figure~\ref{fig2}b presents a grayscale plot of the measured source-drain current $I_{\text{sd}}$ as a function of magnetic field $B$ and plunger gate voltage $V_{\text{PG}}$, acquired at $V_{\text{sd}}$ = 1.2~mV. As $B$ increases from zero, current peaks corresponding to the two spin states of a QD level, along the upper and lower boundaries of a Coulomb blockade diamond with an odd electron occupancy ($N$), split, while those with an even $N$ shift closer together. This is because, for odd $N$, the two spin states derive from the same quantum level and split via the Zeeman effect, while for even $N$, they belong to two adjacent quantum levels. Thus, for an odd-$N$ Coulomb blockade region, the single-particle addition energy evolves as $\Delta \mu(B) = \Delta \mu(0) + \Delta \varepsilon_{n}(B)$, where $\Delta \mu(0)$ is the zero-field single-particle addition energy and $\Delta \varepsilon_{n}(B) = |g_{n}^{*}|\mu_{B}B$ (with $\mu_{B}$ the Bohr magneton) is the Zeeman splitting energy of the $n$-th single-particle level [$n=(N+1)/2$]. Within the linear $B$-dependence region ($B \le$ 1~T), we extract $g$-factors for different quantum levels using the previously determined lever arm and obtain $|g_{1}^{*}|$ = 32, $|g_{2}^{*}|$ = 40, and $|g_{3}^{*}|$ = 55. These values are comparable to those of InSb bulk electrons but exhibit fluctuations, which are in clear agreement with observations in InSb nanowire QDs~\cite{nilsson2009giant}.

\par In addition to the level-dependent $g$-factors, Fig.~\ref{fig2}b reveals a pronounced conductance ridge in the $N=2n+5$ Coulomb blockade region at zero $B$. This ridge occurs at the spin degeneracy point of the $(n+3)$-th QD level and represents a hallmark of the spin-1/2 Kondo effect, which will be discussed in detail later. Furthermore, in the $N=2n+6$ Coulomb blockade region, we observe another weaker, but discernible, enhancement in current at $B \approx 1.2$ T (see Supporting Information for further details). This feature signifies integer-spin Kondo-like correlations, which typically emerge at the degeneracy of two levels of opposite spins and are accompanied by a transition from an $S = 0$ spin-singlet state to an $S = 1$ spin-triplet state as the magnetic field increases~\cite{nilsson2010correlation}.

\par Figure~\ref{fig2}c shows the evolution of two differential conductance peaks, along the red line in Fig.~\ref{fig2}a, as a function of magnetic field. Along this linecut, the electron number in the QD remains unchanged and the two peaks correspond to electron tunneling through the ground and first excited states of the $(2n + 1)$-electron QD. As the magnetic field increases, an anticrossing develops between the spin-down ground state and the spin-up excited state, arising from spin–orbit level mixing. Fitting the peak positions near the anticrossing with a two-level perturbative model yields a spin–orbit splitting of $\Delta_{\text{SO}} = 186~\mu\text{eV}$ (see Supporting Information for details). This value is much smaller than the average level spacing of the InSb nanosheet QD ($\Delta E = 2.4~\text{meV}$), but could result in level-to-level fluctuations in the $g$ factor~\cite{matveev2000g}.

\par
\begin{figure*}
\includegraphics[width=0.5\linewidth]{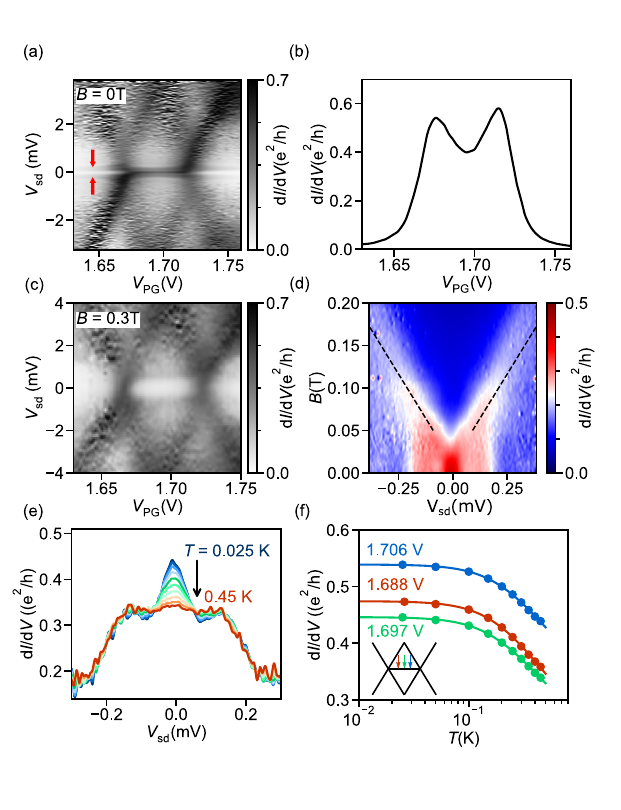}
\caption{
(a) Zoom-in of the $N=2n+5$ charge stability diamond in Figure 2(a) at zero magnetic field, which clearly reveals the spin-1/2 Kondo effect. (b) Kondo ridge at zero bias corresponding to (a). (c) Corresponding charge stability diamond measured at an in-plane magnetic field $B$ = 0.3~T, taken with respect to (a). (d) Differential conductance evolution in the Kondo regime as a function of an in-plane magnetic field  at $V_{\text{PG}}$ = 1.71~V. The dashed line represents a linear fit to the maxima of the differential conductance for $B > 0.05$ T, which is used to extract the effective $g$-factor. (e) Kondo peak evolution as a function of temperature at $V_{\text{PG}}$ = 1.697~V. (f) Kondo peak conductance value as a function of temperature $T$ for different plunger gate voltages within the same Coulomb diamond. Solid circles represent the experimental data, while solid lines indicate the corresponding fits, yielding fitting Kondo temperatures of $T_{K} = 1.04$, $0.94$, and $1.06$ K for $V_{\text{PG}} = 1.688$, $1.697$, and $1.706$ V, respectively. The bottom-left schematic inset indicates the corresponding positions of the three plunger gate voltages within the Coulomb diamond. 
}
\label{fig3}
\end{figure*}

 Next, we investigate the Kondo effect in the superconductor-coupled InSb nanosheet QD. Figure~\ref{fig3}a shows a zoomed-in view of the region near the $N=2n+5$ Coulomb diamond illustrated in Fig.~\ref{fig2}a. Within an even-occupancy Coulomb diamond region (total spin $S=0$), two horizontal differential conductance features are observed at $V_{\mathrm{sd}} \approx \pm 0.2$ mV, as indicated by the arrows in Fig.~\ref{fig3}a. These features correspond to a superconducting gap, signaling the onset of direct quasiparticle tunneling across the junction. In the $N=2n+5$ Coulomb diamond region ($S=1/2$), a zero-bias conductance ridge is observed on top of a broader background conductance. This background originates from proximity-induced subgap transport at zero magnetic field, with ABSs and/or multiple Andreev reflection processes contributing to a finite conductance. As the magnetic field increases and superconductivity is suppressed, this background conductance disappears, as shown in Fig.~\ref{fig3}c. Along the plunger gate voltage axis, this ridge exhibits a characteristic valley or double-peak structure, with the peaks occurring near the charge-degeneracy points (Fig.~\ref{fig3}b). Unlike the nearly suppressed conductance valley in the even-occupancy regime, the valley in the $S=1/2$ region is strongly enhanced (the spin-1/2 Kondo effect). 

\par Figures~\ref{fig3}c and \ref{fig3}d present the response of the Kondo ridge to an in-plane magnetic field. Upon applying an in-plane field of $B=0.3$ T, the superconducting gap previously observed in Fig.~\ref{fig3}a is fully suppressed, thereby restoring the standard Coulomb diamond pattern. In the $2n+5$ Coulomb blockade region, the zero-bias Kondo ridge vanishes and splits into two horizontal differential conductance lines---hallmarks of Kondo splitting. Figure~\ref{fig3}d shows the magnetic-field evolution of the Kondo resonance splitting. From the linear relation $eV_{\mathrm{sd}}=\pm g\mu_{B}B$, we extract an effective $g$-factor of 41 for this quantum level. This value is in reasonable agreement with that determined from Fig.~\ref{fig2}b. 

\par Figure~\ref{fig3}e presents the temperature dependence of the Kondo peak at $V_{\text{PG}}$ = 1.697~V, i.e., at the center of the odd-occupancy diamond. As temperature increases, the Kondo resonance is gradually suppressed and becomes undetectable at $T=0.45$ K. For $T<0.45$ K, the zero-bias conductance exhibits a logarithmic increase (green line in Fig.~\ref{fig3}f) and saturates at low temperatures. This temperature dependence is well described by the empirical scaling function $G(T) = G_{0}/[1+(2^{1/s}-1)(T/T_{K})^2]^{s}$, where $G_{0}$ is the maximum conductance, $T_{K}$ is the Kondo temperature, and $s=0.22$ for a spin-1/2 system~\cite{PhysRevLett.81.5225}. Best fitting to the data yields $T_{K}=0.94$ K. We note that the validity of this empirical formula relies on the superconducting gap remaining essentially unaffected over the experimental temperature range ($T<0.45$ K). Furthermore, temperature-dependent measurements of Kondo peaks were performed at $V_{\text{PG}}$ = 1.688 and 1.706~V within the same diamond. These measurements, analyzed by fitting them to the above empirical scaling function, yield $T_{K}$ = 1.04~K at $V_{\text{PG}}$ = 1.688~V and $T_{K}$ = 1.06~K at $V_{\text{PG}}$ = 1.706~V, revealing higher $T_{K}$ values near the charge-degeneracy points as expected.

\par
\begin{figure*}
\includegraphics[width=1\linewidth]{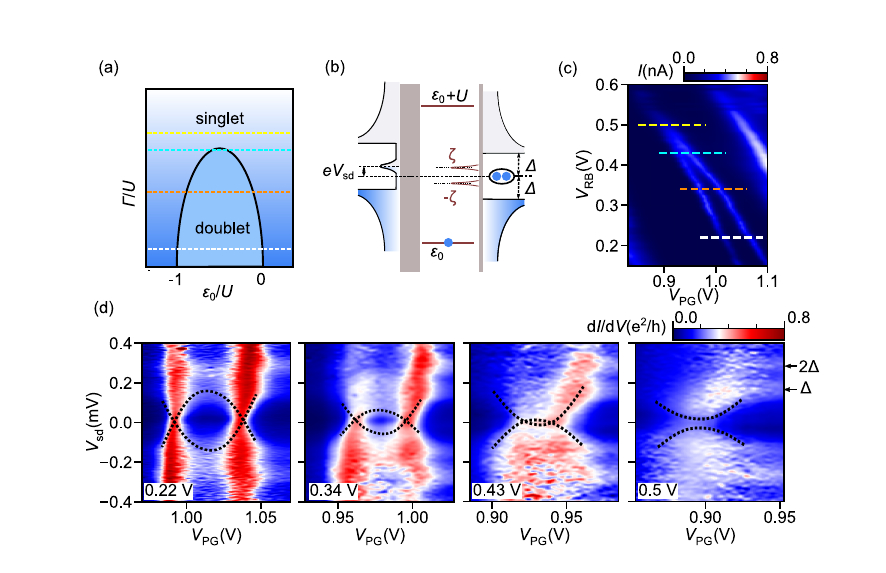}
\caption{
(a) Phase diagram of superconductor-coupled QD. (b) Schematic of the ABS spectroscopy measurements. A soft induced gap in the left probe is taken into account, resulting in a finite quasiparticle density of states near the Fermi level. (c) Source-drain current as a function of plunger gate voltage $V_{\text{PG}}$ and right barrier gate voltage $V_{\text{RB}}$ at a small bias $V_{\text{sd}} = 17.5~\mu V$, with $V_{\text{LB}}$ fixed at -0.2~V. (d) Voltage bias spectroscopy along four different $V_{\text{RB}}$ cuts in (c). The black dashed curves are guides for the crossing behavior of two ABSs resonances. 
}
\label{fig4}
\end{figure*}

 We next investigate singlet-doublet phase transitions in the nanosheet QD-based Josephson junctions. Previous studies have shown that tuning the electron occupancy to an odd value in a superconductor-coupled QD yields two distinct ground states: a spin doublet ($S=1/2$, fermionic odd parity) with two spin degenerate states of $\ket{\uparrow}$ and $\ket{\downarrow}$ ABSs and a spin singlet ($S=0$, fermionic even parity) with $\ket{S}=u\ket{0}-v\ket{\uparrow \downarrow}$~\cite{PhysRevLett.82.2788,PhysRevB.68.035105,PhysRevB.79.224521,lee2014spin}. The latter arises from a Bogoliubov-type superposition of the QD's empty state $\ket{0}$ and two-electron state $\ket{\uparrow \downarrow}$. Four key energy scales govern the competition between these states: the charging energy U, superconductor-QD coupling strength $\Gamma$, superconducting gap $\Delta$, and QD energy level $\varepsilon_{0}$ relative to the electrochemical potential in the superconducting leads. The corresponding phase diagram is shown in Fig.~\ref{fig4}a. For $\varepsilon_{0} > 0$ and $\varepsilon_{0} < -U$, the QD is occupied by 0 and 2 electrons, respectively. A singlet–doublet quantum phase transition is expected only in the regime $-1 < \varepsilon_{0}/U < 0$, where the QD hosts a single (odd) electron. The phase transition point is manifested by a crossing of Andreev bound states at zero energy~\cite{lee2014spin}. As $\Gamma$ increases, this crossing eventually disappears. As demonstrated in previous studies on nanowire systems, such transitions can be effectively probed via voltage bias spectroscopy measurements~\cite{PhysRevLett.104.076805,lee2014spin}. 

\par The theoretical schematic of the ABS spectroscopy measurements is illustrated in Fig.~\ref{fig4}b. A positive voltage applied to $V_{\text{RB}}$ induces ABSs in the QD through strong coupling to the right SC lead, giving rise to Andreev levels $\pm\zeta$. The left SC lead acts as a tunneling probe, with $V_{\text{LB}}$ fixed near pinch-off at $-0.2$~V. When the applied bias $V_{sd}$ aligns the chemical potential of the probe with an Andreev level in the QD ($eV_{sd}=\pm\zeta$), resonance transport occurs, resulting in enhanced current $I$ or differential conductance $d$I /$d$V. When the probe bias is set close to zero, a resonant current is expected as the Andreev level crosses zero energy. Figure~\ref{fig4}c shows the charge stability diagram of the superconductor-coupled QD measured at a small bias voltage ($V_{\text{sd}} = 17.5~\mu$V). $V_{\text{PG}}$ on the horizontal axis tunes the QD energy level $\varepsilon_0$, while $V_{\text{RB}}$ on the vertical axis controls the coupling strength $\Gamma_{\text{RB}}$. Each tilted resonance line corresponds to transport through Andreev levels near zero energy. As $V_{\text{RB}}$ is increased, the two resonance lines merge around $V_{\text{RB}} \approx 0.45$~V. At higher coupling strengths, the resonances progressively weaken and nearly vanish, indicating the absence of zero-energy crossings in the strong-coupling regime. 

By fixing $V_{\text{RB}}$ at various values, we further measure differential conductance versus $V_{\text{PG}}$ and $V_{\text{sd}}$, enabling voltage bias spectroscopy of ABSs across different coupling regimes. Figure~\ref{fig4}d presents such ABS voltage bias spectroscopy results. Although the induced gap is soft, the key features in the spectroscopy remain clearly visible. At weak coupling ($V_{\text{RB}}$ = 0.22~V), bias resonances form a closed “loop” structure, where the upper half-loop opens downward and the lower upward, and they cross at zero bias near $V_{\text{PG}}$ = 0.985~V and $V_{\text{PG}}$ = 1.035~V. This pattern is typically interpreted as evidence of a singlet–doublet ground-state transition occurring at the loop's nodal points. Increasing $\Gamma_{\text{RB}}$ causes the loop structure to shrink gradually [see, for example, the measurement at $V_{\text{RB}}$ = 0.34~V shown in Fig.~\ref{fig4}d] and at $V_{\text{RB}}$ = 0.43~V, the loop structure becomes nearly indiscernible. A further increase in $\Gamma_{\text{RB}}$ causes the two branches to move apart as seen in the case of $V_{\text{RB}}$ = 0.5~V. Notably, these bias resonances occur within the energy range $0 < |eV_{\text{sd}}| < \Delta$. This behavior is consistent with the presence of a soft-gap probe, where a finite density of states of quasi-particles persists near the Fermi energy in the superconducting lead and they tunnel through the ABSs in the superconductor-coupled QD~\cite{su2018mirage}. 

Finally, we need to point out that soft-gap behavior is commonly observed in hybrid nanowire systems~\cite{PhysRevLett.99.126603,PhysRevLett.104.076805,su2018mirage} and is often associated with nonideal superconductor–semiconductor interfaces~\cite{PhysRevLett.110.186803}. Interface processing steps, such as ammonium sulfide etching and Ar-ion cleaning, may introduce interface disorder in our devices, leading to the formation of soft gaps. In contrast, in situ epitaxial Al contacts have been shown to induce hard superconducting gaps in nanowire and quantum well systems~\cite{chang2015hard,kjaergaard2016quantized}, suggesting a promising direction for future device optimization.

\par In conclusion, we report the realization of a superconductor-coupled QD Josephson junction in a planar InSb nanosheet. From a fabrication perspective, our approach utilizes a bilayer fine-gate structure beneath the InSb nanosheet to define the QD in the nanosheet, and contact the QD by depositing superconductor Ti/Al electrodes on top. This device configuration yields a high-quality QD and enables controllable coupling between the QD and superconductor electrodes. Critically, this process can minimize processing on the nanosheet, thereby avoiding detrimental effects on the fabricated device quality. The fabricated superconductor-coupled InSb QD device has been studied by transport measurements. These measurements have shown that the InSb nanosheet QD exhibits large $g$-factors and strong spin-orbit coupling, on par with those of an InSb nanowire QD, as well as diverse quantum phenomena including the Kondo effect and singlet-doublet quantum phase transitions. These results underscore the pivotal role of sub-gap bound states in the transport properties of mesoscopic Josephson junctions and demonstrate that InSb nanosheets could be used as a new platform for exploring the topological physics of Majorana zero modes.

\section{Supporting Information}
Supporting Information. Detailed descriptions of device fabrication and electrical measurements; raw experimental data; and detailed procedures for extracting the spin–orbit energy.

\section{Acknowledgments}
This work is supported by the NSFC (Grant Nos. 92565304, 92165208, 12004039, 92365103, 92576112, 12374480, 12374459, 61974138 and 92065106). D.P. acknowledges the support from the Youth Innovation Promotion Association of the Chinese Academy of Sciences (Nos. 2017156 and Y2021043). 

\section{Conflict of interests}
The authors declare no conflicts of interest. 

\bibliographystyle{apsrev4-2}
\bibliography{ref.bib}


\newpage

\clearpage
\beginsupplement

\begin {center}
\textbf{\large{
Supporting Information: Spin splitting, Kondo correlation and singlet-doublet quantum phase transition in a superconductor-coupled InSb nanosheet quantum dot
}
}
\end {center}

\section{Device fabrication}

~ ~The fabrication process of quantum dot Josephson junction devices begins with the preparation of bilayer metallic electrostatic gates. First, a 60-nm-thick PMMA A2 resist is spin-coated at 4000 rpm for 1 minute on a highly conductive p-type Si(100) substrate with a 300-nm-thick $\text{SiO}_{2}$ layer, followed by baking at 170°C for 10 minutes. Standard electron-beam lithography (EBL) is then performed on the PMMA at an electron accelerating voltage of 125 kV. The patterned sample is developed in MIPK:IPA (1:3) for 1 minute, followed by immersion in IPA for 30 seconds, and dried with nitrogen before evaporating the Ti/Au plunger gate. After fabricating the plunger gate, the sample is transferred to an atomic layer deposition (ALD) system to grow a 10-nm $\text{Al}_{2}\text{O}_{3}$ layer. The same EBL process is repeated to fabricate the second-layer metallic electrodes and dielectric layers.

After preparation of the bilayer metallic electrostatic gates, InSb nanosheets are transferred onto the bilayer gates using a mechanical probe. To ensure good ohmic contact between the InSb nanosheets and superconducting electrodes, the sample is treated with a 40 $^\circ C$ ammonium polysulfide solution (3 mol/L $(\text{NH}_4)_2\text{S}_{x}$: $\text{H}_2\text{O}$ = 1:400 by volume) for 40 minutes to remove surface oxides. The sample is then quickly transferred to an evaporation chamber. In situ Ar plasma etching (40 s, 3 W, Ar pressure $7\times10^{-4}$ mbar) is performed before the evaporation of Ti/Al (5 nm/100 nm) as superconducting electrodes.

\section{Measurement circuit setup}

~ ~A schematic of the external voltage-biased measurement setup is shown in Fig.~S1. DC signals were generated by a Yokogawa GS200 source, routed through a voltage divider (100:1 or 1000:1 depending on the measurement configuration), and applied to the device source electrode. The resulting drain current was amplified by a low-noise current preamplifier (SR570) and recorded using a digital multimeter (Agilent 34401A). The gate voltages \( V_{\text{PG}}, V_{\text{LB}}, V_{\text{RB}}, V_{\text{CG}}, \) and \( V_{\text{CO}} \) were controlled by a QDevil QDAC.

\begin{figure}[H]\centering
\includegraphics[width=0.6\linewidth]{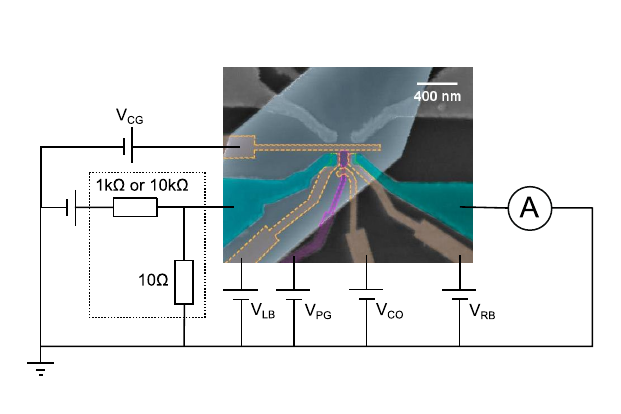}
\caption{\label{fig:s1} 
Schematics of the measurement circuit setup for the voltage-biased configuration. The region enclosed by the dashed line denotes a voltage divider.
}
\end{figure}

\section{Raw data for Figure 3f}

\begin{figure}[H]\centering
\includegraphics[width=0.8\linewidth]{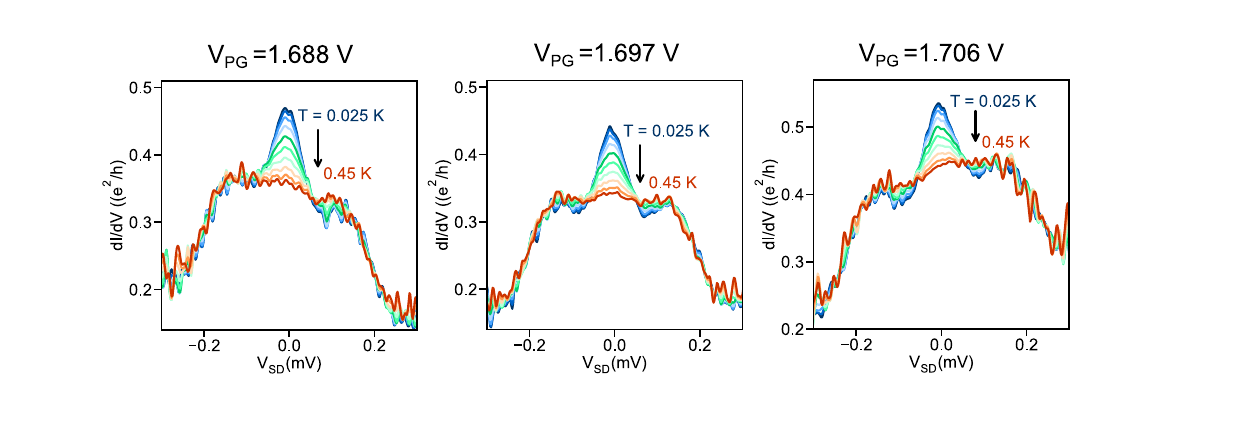}
\caption{\label{fig:s2} 
Temperature dependence of Kondo peaks at three different $V_{\text{PG}}$. These three gate voltages are all located within the same Coulomb diamond: $V_{\text{PG}}$ = 1.697 V is near the diamond’s center, while $V_{\text{PG}}$ = 1.688 V and 1.706 V are closer to the charge-degeneracy points. Schematics of these positions are marked in the inset of Fig. 3f of the main article.
}
\end{figure}

\section{Raw data along the blue cut in the $N=2n+6$ Coulomb blockade region shown in Figure 2b}

\begin{figure}[H]\centering
\includegraphics[width=0.5\linewidth]{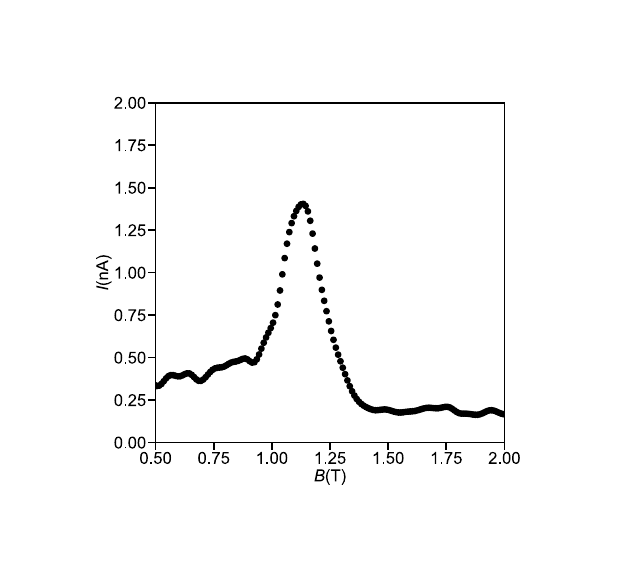}
\caption{\label{fig:s4} 
Source-drain current as a function of magnetic field at $V_{\text{PG}}$ = 1.75~V. We observe a conductance enhancement due to integer-spin Kondo-like correlations.
}
\end{figure}

\section{Extraction of the spin-orbit energy $\Delta_{\text{SO}}$ by fitting the experimental data}

~ ~To extract the spin-orbit energy $\Delta_{\text{SO}}$, we employ a two-level perturbation model with the Hamiltonian:
\[
H = \begin{bmatrix} 
\varepsilon_{n0} + \frac{1}{2} |g_{n0}^{*}| \mu_{\text{B}} B & \Delta_{\text{SO}} \\ 
\Delta_{\text{SO}} & \varepsilon_{n0+1} - \frac{1}{2} |g_{n0+1}^{*}| \mu_{\text{B}} B 
\end{bmatrix},
\]
Here, $\varepsilon_{n0}$ and $\varepsilon_{n0+1}$ are the ground- and the excited-state energy levels, $g_{n0}^{*}$ and $g_{n0+1}^{*}$ are the $g$-factors of the two levels. Fitting the magnetic field dependence of the eigenvalues of this Hamiltonian to the experimental data yields $|g_{n0}^{*}|$ = 50 (ground state), $|g_{n0+1}^{*}|$ = 23 (excited state), and $\Delta_{\text{SO}} = 186\ \mu\text{eV}$. Notably, the spin-orbit energy in the InSb QD is much smaller than the average level spacing ($\Delta E$ = 2.4~meV).

\end{document}